\documentclass{article}
\usepackage{spconf,amsmath,graphicx}
\usepackage{url}

\usepackage{float}

\title{Unsupervised Learning of Audio Perception for Robotic Applications:
\newline
Learning to project data to T-SNE/UMAP Space}
\name{Prateek Verma$^*$ and Kenneth Salisbury \thanks{$^*$This work was done while Prateek Verma was affiliated with the Stanford Computer Science Department.}}
\address{Stanford Artificial Intelligence Laboratory\\Stanford University}
\begin{document}
%
\maketitle
\begin{abstract}

\end{abstract}
Audio perception is a key to solving a variety of problems ranging from acoustic scene analysis, music meta-data extraction, recommendation, synthesis and analysis. It can potentially also augment computers in doing tasks that humans do effortlessly in day-to-day activities. This paper builds upon key ideas to build perception of touch sounds without access to any ground-truth data. We show how we can leverage ideas from classical signal processing to get large amounts of data of any sound of interest with a high precision. These sounds are then used, along with the images to map the sounds to a clustered space of the latent representation of these images. This approach, not only allows us to learn semantic representation of the possible sounds of interest, but also allows association of different modalities to the learned distinctions. The model trained   to map sounds to this clustered representation, gives reasonable performance as opposed to expensive methods collecting a lot of human annotated data. Such approaches can be used to build a state of art perceptual model for \textit{any} sound of interest described using a few signal processing features.
Daisy chaining high precision sound event detectors using signal processing combined with neural architectures and high dimensional clustering of unlabelled data is a vastly powerful idea, and can be explored in a variety of ways in future.\newline
\section{Introduction}
\label{sec:intro}

Audio scene understanding has been a subject that has been studied in depth for the past couple of decades \cite{bregman1994auditory}. In addition to making computers understand human speech, giving them the capabilities to hear and understand everyday sound has enormous applications. 
Giving computers ability to hear, see and speak on par with humans has been a long standing vision of early works in artificial intelligence \cite{papert1966summer}. With the advent of convolutional models early success in vision-based CNN architectures have been translated to audio based research with proportionate gains in performance \cite{hershey2017cnn}. Various architectures from vision and natural processing have been successfully deployed to solve various problems in audio transformations \cite{haque2018conditional}, style transfer \cite{verma2018neural}, speech recognition \cite{chan2015listen},  learning latent representations \cite{haque2019audio}, speech to speech translation \cite{guo2019end}, synthesis \cite{wang2017tacotron}, perception \cite{verma2019learning}. As quoted by a famous researcher Prof. Andrew Zisserman, "Given enough data, and a valid existing one to one mapping from one signal to another, deep neural networks can solve any task irrespective of the domain". Additionally, they are also able to unearth hidden patterns, which are inherently difficult to decipher without access to such trained million parameter black boxes. The authors of \cite{poplin2018prediction} showed that one can get various attributes like sex, age, smoker etc. just from retinal images of eye. 
\newline
One application of learning good perception model has been in robotics in problems such as insertion and grasping. The work by \cite{lee2019making} used shared latent representation across modalities such as vision, touch to solve the problem of interest. In future, one can imagine that these latent based representations will be universal irrespective of the domain to solve a particular task. For robotics, these have already been shown to advance the state of the art in problems with supervised/unsupervised approaches. Our work explores how we can augment similar auditory based representations. Additionally, we also propose to have these models learn from actual unlabelled data present in day to day interactions and vast amounts of data e.g. YouTube to solve a particular perception task viz. Audio perception in our case. The addition of sound in robotics application has been relatively untapped and may likely augment existing vision and touch based sensors in future given similarities in learning algorithms and architectures in these domains \cite{lee2019making}. This work builds one such sub-block namely can we build a perceptual model for any sound of interest for robotic application. Such models have similar architecture as existing one proposed in \cite{lee2019making} and will help in learning latent representations for a particular scenario of interest e.g. scratching, rubbing, tapping. 
There has been works in exploring how can we build perceptual models for sound of interest by collecting vast amounts of data \cite{hershey2017cnn}. The work done by \cite{owens2016visually} tapped a variety of objects to predict the sound and material properties using touch sound as shown in Figure 1.

\begin{figure}[H]

  \centerline{\includegraphics[width=8.5cm]{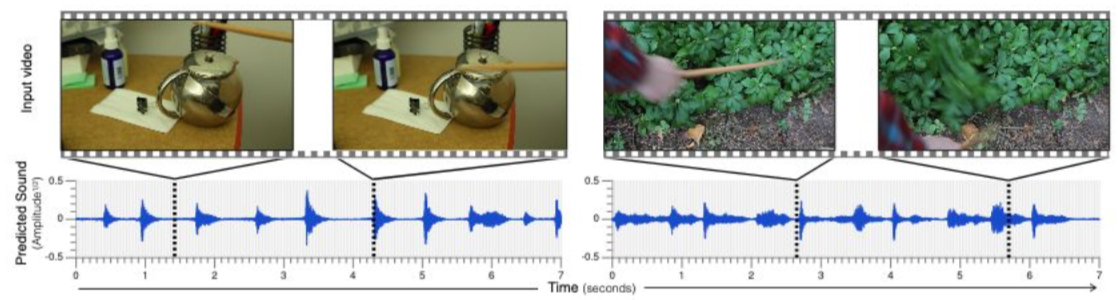}}
\caption{Diagram from \cite{owens2016visually} indicating creation of a dataset to record sounds while recording video of impact sound caused by beating a stick to different objects. This approach is not scalable. }
\label{fig:mit}
\end{figure}
However, such data collection is expensive, uses a lot of resources, and cannot be scaled for a wider context and interactions.  We in this work propose a method of unsupervised learning a perceptual model for signal of interest, viz. touch sounds by daisy chaining signal processing to get sounds of interest with a high precision. We then use convolutional architectures to learn how to project sounds to a Uniform Manifold Approximation space \cite{mcinnes2018umap} of the corresponding images. This bypasses the need of any labelled data. 
\par 
The flow of paper is as follows: Section 2 describes the data-set used for the current paper. It is followed by the methodology in Section 3 which details various signal processing techniques to build a high precision detector. It also elucidates the need for building the deep learning model and a clustering algorithm used for the contents of images. This is followed by relation to prior work and references.

\section{Data-set}
We took as input 5000 YouTube videos from a variety of real world scenarios. A subset of AudioSet \cite{gemmeke2017audio} corpus was chosen, which consisted of sounds like chop, tap, rub, slap, hammer etc.
Due to building a high precision detector, we had a total of 3000 training examples, with some chosen for validation and testing. Given the availability of such datasets like YouTube-8M \cite{abu2016youtube} and AudioSet, in future, this can be scaled across a large variety of scenarios. We do not work with the availability of pretrained embeddings, as they are only trained for a particular application, and are not trained to infer fine-grained distinctions i.e. in our case various types of sounds of touch. The videos in addition to their corresponding audio were downloaded, down-sampled to 16kHz. Despite having different frame rates and resolution, were all stored by down-sampling/up-sampling to the resolution of (224,224,3) to make the images consistent with that of pretrained architectures trained on Image-Net.

\section{Methodology}
This section explains the ideas and method that we have discussed so far in order to build and train a neural network for building a perception model for the sound of interest.
\subsection{Signal Processing to Understand Sounds} A rich literature of hand crafted features was developed to capture various time and frequency domain characteristics before the advent of deep learning \cite{audiofeatures}. These features, even though have been collectively surpassed with the power and expressivity of deep neural networks having millions of parameters can still be used to richly explore certain characteristics of the signal of interest. We explain a few of these, in order to better motivate and emphasize their use. Some of the features that have been used both at a micro and macro level are zero crossings, spectral centroid, spectral flatness, onset strength, energy contour to name a few. Figure 2 explains a method of computing onset strength at each of the points in log-magnitude spectogram with the height of the red curve depicting the strength of the onset. We will explore a few of these features as described in the next section.

\begin{figure}[H]
  \centerline{\includegraphics[width=8.5cm]{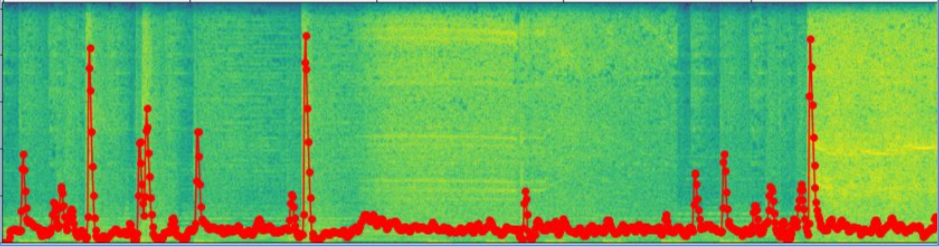}}
\caption{Spectral onsets computed for each of the time points in the spectogram. They have a strong correlation with impact and touch sounds}
\label{fig:fig2}
\end{figure}

\subsection{Detecting Sounds with High Precision} 
From the rich variety of features available to us, we explore how to characterize touch/impact sounds for the current work. 
Consider a time domain signal $x[k]$ and its corresponding spectogram denoted as $X[k,m]$ where $k$ denotes the $k^{th}$ frequency bin computed at time instance $m$ typically every 10ms. We denote the following features as follows: Energy contour, $E[n]$ is defined as, 

\begin{equation}
 E[n] = \sum_{k=1}^{N} |X[k,n]|
\end{equation}

Spectral centroid, as the name suggests is the center of mass of each of the spectral slices $X[:,n]$ and computed as the weighted average of the spectral weights over the spectral bin indices. Figure 3 describes the spectral centroid and the energy onsets of a setting where some utensils were moved and dropped. Spectral flatness $SF[n]$, is defined as,

\begin{equation}
 SF[n] = \frac{\sqrt[n]{\prod_{k=1}^{N}|X[k,n]|}}{\frac{1}{N}\sum_{k=1}^{N} |X[k,n]|} 
\end{equation}
Spectral flatness is a measure of the impulsiveness of the sound or how the spectral spread of the spectral slice is across the frequency bins. Additionally, we also describe the spectral attributes or changes in terms of onsets.
\begin{figure}[H]
  \centerline{\includegraphics[width=8.5cm,height=3cm]{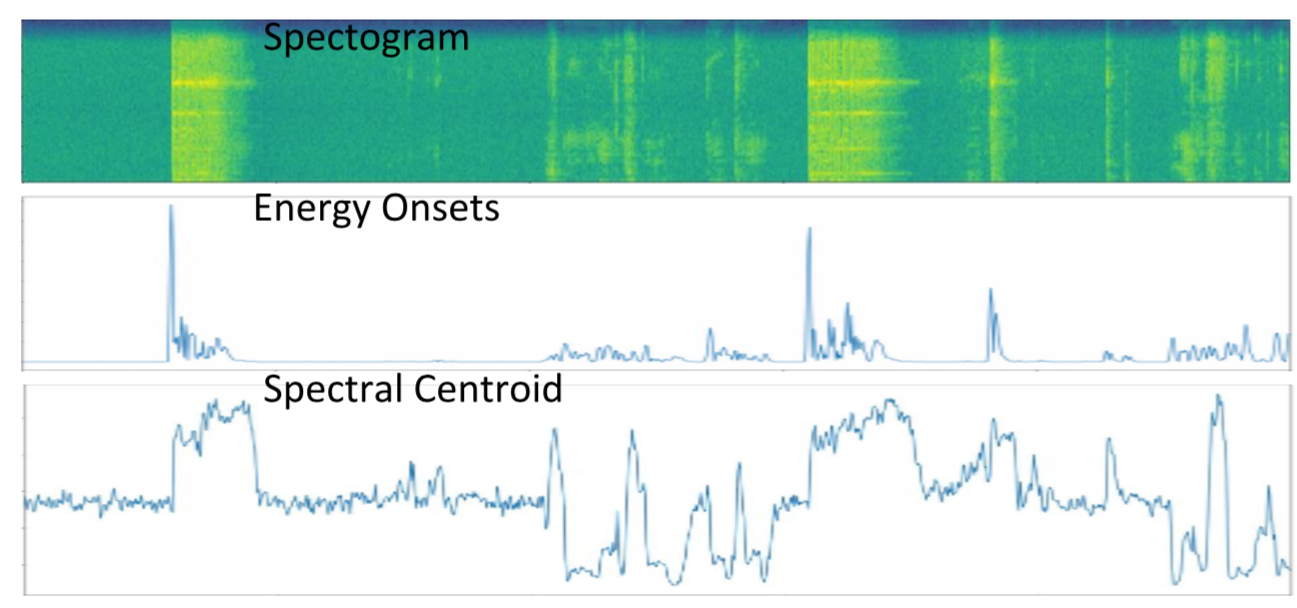}}
\caption{Example of a spectogram (top), energy onsets (middle) and spectral centroid of the spectogram slice (bottom) }
\label{fig:fig1}
\end{figure}

  A onset function $O[n]$, computed every 10ms is, 

\begin{equation}
O[n] = \sum_{k=1}^{N} \{H(|X[k,n]|-|X[k,n-1]|)\}
\end{equation}

where the half wave rectification function is defined as $H(x) = \frac{|x|+x}{2}$. Another attribute, the number of zero crossings is defined as the number of times signal crosses the mean value in the waveform domain, quite often 0. Periodic signals will have a small average zero crossings in a frame whereas noisy/impulsive signals will have larger values. The touch sounds can be characterized using the following observations i) they consist of relative peaks in the energy envelop irrespective of the background noise present if any, ii) they usually are a wide-band event having a wide spectral spread as they are often associated with impulsive events iii) they exhibit peaks in spectral onsets. A smoothed energy contour is computed, and in addition to peaks, we also compare the depth of the peaks to correlate it with the impact/touch sound. In addition, we search for peaks in spectral flatness above a threshold and relative onset strength. The parameters were chosen by hand, in order to build the models with a high precision and relatively poor recall. Figure 4 describes a energy contour with the choice of the points chosen in red, in building a high precision detector.

\subsection{Building Deep Learning Models}
There has been a rich variety of work in using convolutional and recurrent models to build audio understanding systems \cite{hershey2017cnn} \cite{verma2019neuralogram}. For most of the problems in classification, if there is availability of large amounts of data, these million parameter models often are able to learn these mappings from one domain of interest to another, in this case, one hot encoding of the category of the sound. Often, availability of multi-modal streams of data can be helpful in learning one domain from another. Sound-Net \cite{aytar2016soundnet} leveraged a state of the art image understanding system to understand the contents of the audio signal, with a mean squared error loss to train a convolutional net to map wave-forms to latent space of images. 

\begin{figure}[H]
  \centerline{\includegraphics[width=8.5cm, height=3cm]{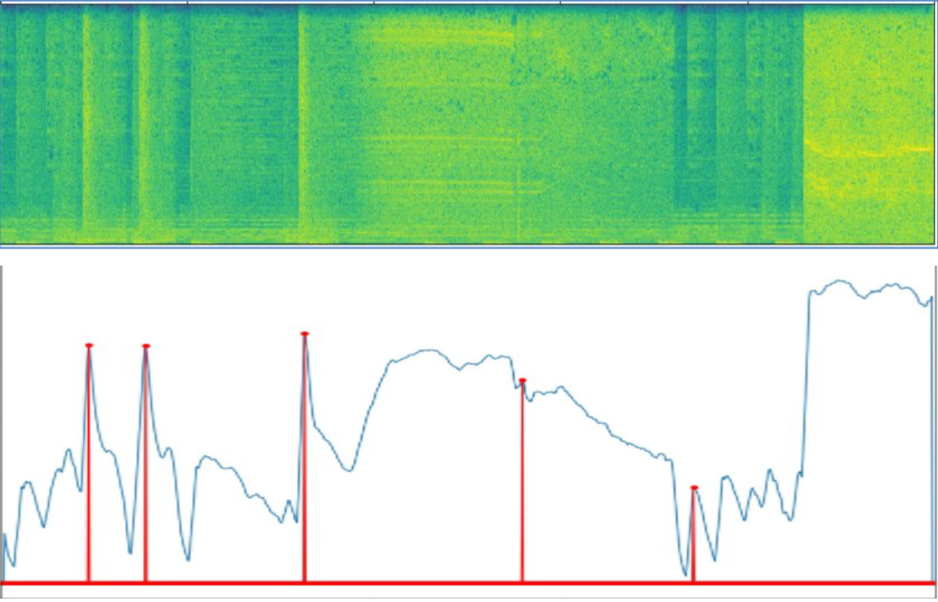}}
\caption{Original Spectogram (top) along with the energy contour and peaks chosen in red for training our system.}
\label{fig:fig1}
\end{figure}
However, such approaches do not work for the current problem of interest in building perceptual models for touch/impact sounds. These sounds are often too small in duration roughly around 40-100ms. Trying to map these sounds to the latent spaces, often ends up learning other sounds than the touch sounds due to data imbalancing problems. Given a spectra of say 10sec, such sounds will only account for <1 \% of the frames and the minimization criteria will only optimize the errors present in other 99 \% of the data points most of the times. The data imbalance problem in machine learning is still an active area of research, and people often mitigate such issues with sub-sampling, loss functions. However, trying to build such models in unsupervised setting is difficult to do as we are not aware of the position of sampling.
Additionally, the latent space characterizes the salient contents of the image, and not necessarily the subtle characteristics present for distinguishing the sounds. Consider the examples as shown in the Figure 4.

\begin{figure}[htb]

  \centerline{\includegraphics[width=8.5cm]{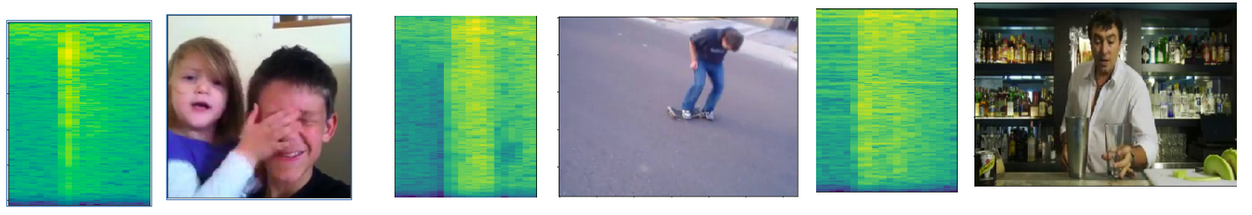}}

%

\caption{Example of the sounds and the corresponding images. The representation is a log magnitude linear spectogram with 129 bins and the duration of 100ms. Notice how we achieve rich variety along with the images }
\label{fig:fig1}
\end{figure}

For all the three of the cases, the image embeddings will give a salient weightage of humans/person but the corresponding sound describe three distinct events namely slapping, skateboarding and placing a glass on table. We can argue that we can have additional constraints like incorporation of a multi-category loss \cite{aytar2016soundnet}, but it would not help us in understanding the contents and its correspondence to the other images present in the dataset. Instead of focusing on having diverse embeddings as target and addition of metadata information, we focus purely on a approach based on clustering. This will also mitigate the need to predict the non-salient parts of the images e.g. person, drinks in the background etc in the above example. One point to note is that a siamese architecture, which associates similarities between images and audio has been shown to produce meaningful results in learning latent representations for both the modalities \cite{arandjelovic2017look}. This paper explores yet another way to learn unsupervised latent representations for the problem of interest, in this case audio perception of touch sounds. 
\subsection{Uniform Manifold Approximation}
Uniform Manifold Approximation and Projection \cite{mcinnes2018umap} is currently a state of the art clustering and dimension reduction algorithm that uses graph theory, fuzzy sets and topology. The embedding is found by searching for a low dimensional projection of the data that has the closest possible equivalent fuzzy topological structure. Across several datasets, it has been shown to retain the overall global structure present in the input signal of interest which is hard to capture with techniques such as PCA or TSNE \cite{maaten2008visualizing}.It also allows us to adjust the parameter space of UMAP more intuitively than TSNE, resulting in better visualization and clustering in a semantically meaningful manner. UMAP constructs a high dimensional graph representation of the data then optimizes a low-dimensional graph to be as structurally similar as possible. 
\par
We tuned two parameters namely number of neighbours as well as minimum distance  which adjusts local vs global structure and the manner of clumping together of data. By adjusting these parameters, we can see we achieve much better separation and clustering of ~ 3000 training images. UMAP embeddings also remove all the portions of the latent space (or embedding coordinates) that are not relevant to the problem of interest, and cluster images based upon its close similarity with other possible images present in the data-set. As we can see from doing similar ideas in images, e.g. T-SNE \cite{maaten2008visualizing} on imagenet, it somehow encodes the images into a space that has similar images in its vicinity and unrelated images further away from it. We compute UMAP representations for the latent codes learned from VGG-19 architectures. VGG-19 pretrained architecture was chosen as a design choice, and can be easily replaced by other state of the art pretrained models for getting better latent representations. Figure 6 shows the difference in the clustering performances of these two algorithms, and see how UMAP achieves better performance than T-SNE. In order to assess the quality of clustering, we sample images from UMAP space and its vicinity. We find that the images are  clustered according to the type of impact seen in the images in addition to the overall scene. We see from Figure 7  that we can get drill sound, chiseling of wood, lab equipment based sound and dropping of a log of wood.

\subsection{Mapping sounds to UMAP space}
In order to learn the latent representations of a sounds, we use the position in the UMAP as a supervision. A 5 layer CNN model similar to \cite{arandjelovic2017look} with 3x3 filters and 2x2 pooling was used, with Euclidean loss as the error criterion from the prediction points to the actual position in the UMAP space. The resolution of the spectogram was as follows: 10ms hop, 16kHz sampling rate, with 30ms window size and 512 pt FFT, with a total of 200ms of input in duration. This gives us a spectogram sized 257x20. The points in UMAP space will help guide us to the right cluster. Given a lot of data, it can help us perhaps in learning even finer grained distinctions present in the data. 

\begin{figure}[H]
  \centerline{\includegraphics[width=8.5cm]{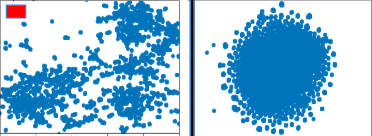}}
\caption{UMAP embeddings of 3000 images on left vs TSNE points on the right. Notice how the space is nicely clustered. The red patch shows the error in the test cases in mapping unknown sound to UMAP space}
\label{fig:fig1}
\end{figure}

However, for most of the coarse settings such as "whether this was a wooden impact or not", we can already build state of the art models from access to a small number of such sounds. In order to evaluate the performance of our system, we hold out a portion of the test sounds, while \textit{not} keeping the corresponding images out of the training set.  The performance of the system is depicted in Fig 6, where the red square gives the average error in the x-y coordinates in projecting a test set of sounds into UMAP coordinates.

\begin{figure}[H]
  \centerline{\includegraphics[width=8.5cm]{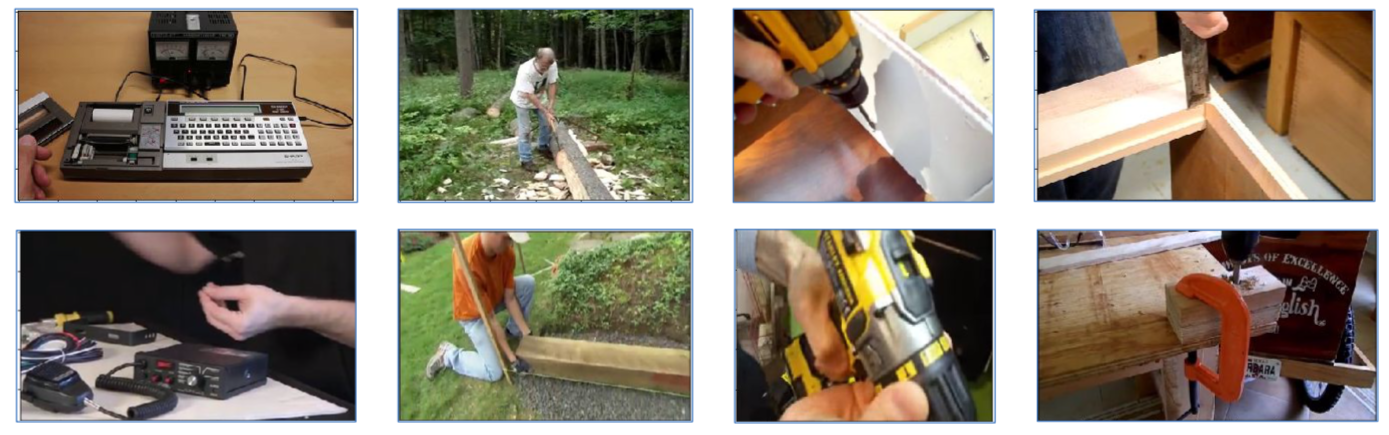}}
\caption{Reference image in UMAP space (top), and sampled retrieved image (bottom)}
\label{fig:fig1}
\end{figure}


\section{RELATION TO PRIOR WORK}
\label{sec:prior}
This section emphasizes our contributions related to prior work in the field. We believe that there are several ideas that can be exploited in a variety of situations and domains. Mapping to UMAP space rather than labels or embeddings is a vastly untapped idea and can be used in several setups. Additionally, combining signal processing with deep learning has been done to some extent in the past \cite{verma2019learning}, but this work utilizes signal processing to collect cheap, high precision data from unlabelled sources. Such ideas can be used in future to get any sound of interest and get corresponding modalities e.g. vision. This work also motivates, as to how when the signal of interest is small enough, how the traditional mappings to a latent representation would fail due to imbalance issues (between the number of points of signal of interest) while minimizing the euclidean distances between the latent representation and the embeddings learned by a convolutional architecture.


\bibliographystyle{IEEEbib}

\bibliography{strings}

\begin{thebibliography}{10}

\bibitem{bregman1994auditory}
Albert~S Bregman,
\newblock {\em Auditory scene analysis: The perceptual organization of sound},
\newblock 1994.

\bibitem{papert1966summer}
Seymour~A Papert,
\newblock ``The summer vision project,''
\newblock 1966.

\bibitem{hershey2017cnn}
Shawn Hershey, Sourish Chaudhuri, Daniel~PW Ellis, Jort~F Gemmeke, Aren Jansen,
  R~Channing Moore, Manoj Plakal, Devin Platt, Rif~A Saurous, Bryan Seybold,
  et~al.,
\newblock ``Cnn architectures for large-scale audio classification,''
\newblock in {\em 2017 ieee international conference on acoustics, speech and
  signal processing (icassp)}. IEEE, 2017, pp. 131--135.

\bibitem{haque2018conditional}
Albert Haque, Michelle Guo, and Prateek Verma,
\newblock ``Conditional end-to-end audio transforms,''
\newblock {\em arXiv preprint arXiv:1804.00047}, 2018.

\bibitem{verma2018neural}
Prateek Verma and Julius~O Smith,
\newblock ``Neural style transfer for audio spectograms,''
\newblock {\em arXiv preprint arXiv:1801.01589}, 2018.

\bibitem{chan2015listen}
William Chan, Navdeep Jaitly, Quoc~V Le, and Oriol Vinyals,
\newblock ``Listen, attend and spell,''
\newblock {\em arXiv preprint arXiv:1508.01211}, 2015.

\bibitem{haque2019audio}
Albert Haque, Michelle Guo, Prateek Verma, and Li~Fei-Fei,
\newblock ``Audio-linguistic embeddings for spoken sentences,''
\newblock in {\em ICASSP 2019-2019 IEEE International Conference on Acoustics,
  Speech and Signal Processing (ICASSP)}. IEEE, 2019, pp. 7355--7359.

\bibitem{guo2019end}
Michelle Guo, Albert Haque, and Prateek Verma,
\newblock ``End-to-end spoken language translation,''
\newblock {\em arXiv preprint arXiv:1904.10760}, 2019.

\bibitem{wang2017tacotron}
Yuxuan Wang, RJ~Skerry-Ryan, Daisy Stanton, Yonghui Wu, Ron~J Weiss, Navdeep
  Jaitly, Zongheng Yang, Ying Xiao, Zhifeng Chen, Samy Bengio, et~al.,
\newblock ``Tacotron: Towards end-to-end speech synthesis,''
\newblock {\em arXiv preprint arXiv:1703.10135}, 2017.

\bibitem{verma2019learning}
Prateek Verma and Jonathan Berger,
\newblock ``Learning to model aspects of hearing perception using neural loss
  functions,''
\newblock {\em arXiv preprint arXiv:1912.05683}, 2019.

\bibitem{poplin2018prediction}
Ryan Poplin, Avinash~V Varadarajan, Katy Blumer, Yun Liu, Michael~V McConnell,
  Greg~S Corrado, Lily Peng, and Dale~R Webster,
\newblock ``Prediction of cardiovascular risk factors from retinal fundus
  photographs via deep learning,''
\newblock {\em Nature Biomedical Engineering}, vol. 2, no. 3, pp. 158, 2018.

\bibitem{lee2019making}
Michelle~A Lee, Yuke Zhu, Krishnan Srinivasan, Parth Shah, Silvio Savarese,
  Li~Fei-Fei, Animesh Garg, and Jeannette Bohg,
\newblock ``Making sense of vision and touch: Self-supervised learning of
  multimodal representations for contact-rich tasks,''
\newblock in {\em 2019 International Conference on Robotics and Automation
  (ICRA)}. IEEE, 2019, pp. 8943--8950.

\bibitem{owens2016visually}
Andrew Owens, Phillip Isola, Josh McDermott, Antonio Torralba, Edward~H
  Adelson, and William~T Freeman,
\newblock ``Visually indicated sounds,''
\newblock in {\em Proceedings of the IEEE conference on computer vision and
  pattern recognition}, 2016, pp. 2405--2413.

\bibitem{mcinnes2018umap}
Leland McInnes, John Healy, and James Melville,
\newblock ``Umap: Uniform manifold approximation and projection for dimension
  reduction,''
\newblock {\em arXiv preprint arXiv:1802.03426}, 2018.

\bibitem{gemmeke2017audio}
Jort~F Gemmeke, Daniel~PW Ellis, Dylan Freedman, Aren Jansen, Wade Lawrence,
  R~Channing Moore, Manoj Plakal, and Marvin Ritter,
\newblock ``Audio set: An ontology and human-labeled dataset for audio
  events,''
\newblock in {\em 2017 IEEE International Conference on Acoustics, Speech and
  Signal Processing (ICASSP)}. IEEE, 2017, pp. 776--780.

\bibitem{abu2016youtube}
Sami Abu-El-Haija, Nisarg Kothari, Joonseok Lee, Paul Natsev, George Toderici,
  Balakrishnan Varadarajan, and Sudheendra Vijayanarasimhan,
\newblock ``Youtube-8m: A large-scale video classification benchmark,''
\newblock {\em arXiv preprint arXiv:1609.08675}, 2016.

\bibitem{audiofeatures}
Geoffroy Peters,
\newblock {\em A Large Set of Audio Features}, 2003 (accessed February 8,
  2020).

\bibitem{verma2019neuralogram}
Prateek Verma, Chris Chafe, and Jonathan Berger,
\newblock ``Neuralogram: A deep neural network based representation for audio
  signals,''
\newblock {\em arXiv preprint arXiv:1904.05073}, 2019.

\bibitem{aytar2016soundnet}
Yusuf Aytar, Carl Vondrick, and Antonio Torralba,
\newblock ``Soundnet: Learning sound representations from unlabeled video,''
\newblock in {\em Advances in neural information processing systems}, 2016, pp.
  892--900.

\bibitem{arandjelovic2017look}
Relja Arandjelovic and Andrew Zisserman,
\newblock ``Look, listen and learn,''
\newblock in {\em Proceedings of the IEEE International Conference on Computer
  Vision}, 2017, pp. 609--617.

\bibitem{maaten2008visualizing}
Laurens van~der Maaten and Geoffrey Hinton,
\newblock ``Visualizing data using t-sne,''
\newblock {\em Journal of machine learning research}, vol. 9, no. Nov, pp.
  2579--2605, 2008.

\end{thebibliography}

\end{document}